\documentclass[twocolumn,showpacs,preprintnumbers,superscriptaddress,amsmath,amssymb,prl]{revtex4}

\usepackage{graphicx}
\usepackage{dcolumn}
\usepackage{bm}

\begin{document}

\preprint{APS/123-QED}

\title{Direct observation of non-classical photon statistics
    in parametric downconversion}
\author{Edo Waks}
\affiliation{Quantum Entanglement Project, ICORP, JST, E.L.
Ginzton Laboratories, Stanford University, Stanford, CA 94305 }
\author{Eleni Diamanti}
\affiliation{Quantum Entanglement Project, ICORP, JST, E.L.
Ginzton Laboratories, Stanford University, Stanford, CA 94305 }
\author{Barry C. Sanders}
\affiliation{Quantum Entanglement Project, ICORP, JST, E.L.
Ginzton Laboratories, Stanford University, Stanford, CA 94305 }
\affiliation{Department of Physics, Macquarie University,
    Sydney, New South Wales 2109, Australia}
\affiliation{Department of Physics and Astronomy, University of Calgary,
    Alberta T2N 1N4, Canada}
\author{Stephen D. Bartlett}
\affiliation{Department of Physics, Macquarie University, Sydney, New
  South Wales 2109, Australia}
\author{Yoshihisa Yamamoto\cite{byline}}
\affiliation{Quantum Entanglement Project, ICORP, JST, E.L.
Ginzton Laboratories, Stanford University, Stanford, CA 94305 }

\date{17 July 2003}

\newcommand{\avg}[1]{\langle #1 \rangle}
\def\Re{\mbox{Re}}
\def\Im{\mbox{Im}}
\newcommand{\TR}[1]{\mbox{Tr}\left\{ #1 \right\}}
\newcommand{\TRA}[1]{\mbox{Tr}_a\left\{ #1 \right\}}
\newcommand{\TRE}[1]{\mbox{Tr}_e\left\{ #1 \right\}}
\newcommand{\TRAE}[1]{\mbox{Tr}_{ae}\left\{ #1 \right\}}
\def\<#1|{\langle#1|}
\def\|#1>{|#1\rangle}
\def\Trans{\alpha_L}
\def\T2{\alpha_{L/2}}
\def\TM{\alpha_{N}}
\def\nbar{\bar{n}}
\def\adag{\hat{a}^{\dagger}}
\def\bdag{\hat{b}^{\dagger}}
\def\a{\hat{a}}
\def\b{\hat{b}}
\newcommand{\n}{\hat{n}}
\newcommand{\tr}{\mbox{Tr}}
\def\bk<#1|#2>{\left\langle\vphantom{#1|#2}#1\right|%
\left.\vphantom{#1|#2}#2\right\rangle}
\def\kb|#1><#2|{\left|\vphantom{#1|#2}#1\right\rangle%
\left\langle\vphantom{#1|#2}#2\right|}
\def\aux|#1><#2|{\left|\vphantom{#1|#2}#1\right\rangle_{a}%
\left\langle\vphantom{#1|#2}#2\right|}
\def\matrix[#1][#2]{\left[
  \begin{array}{#1}
    #2
  \end{array}  \right]}
\def\g2{g^{(2)}}

\begin{abstract}

We employ a high quantum efficiency photon number counter
to determine the photon number distribution of the output field
from a parametric downconverter. The raw photocount data directly
demonstrates that the source is nonclassical by forty standard
deviations, and correcting for the quantum efficiency yields
a direct observation of oscillations in the photon number distribution.

\end{abstract}

\maketitle

Quantum optics~\cite{Man95,Sch01} is concerned with optical
phenomena that cannot be described by a classical field treatment.
Operationally, photon correlations are measured and
nonclassicality is established by violations of correlation
inequalities imposed by the assumption that the field is
classical. Well known examples are observations of sub-Poissonian
photon number statistics~\cite{Sho83}, photon
antibunching~\cite{Kim77}, and photocurrent fluctuations below the
shot noise level for squeezed light sources~\cite{Slu85}. Although
an equivalence between sub-Poissonian statistics and photon
antibunching has sometimes been assumed, the two phenomena are
distinct revelations of a nonclassical field~\cite{Sin83,Zou90},
and the choice of test for nonclassicality may need to depend on
the source. Both measures are adequate to test nonclassicality for
resonance fluorescence, but antibunching is certainly not
sufficient to identify nonclassicality for two-photon sources such
as an atomic cascade~\cite{Koc67} or parametric downconversion
(PDC)~\cite{Bur70}. Sophisticated demonstrations of
nonclassicality for two-photon sources have been employed such as
showing complementarity~\cite{Gra86} or violating a Bell
inequality~\cite{Asp81}. However, direct measurement of the photon
number distribution may be sufficient to demonstrate
nonclassicality, and such a measurement is appealing because the
nonclassicality criterion is strictly based on the actual
measurement record, inequalities that arise from a classical field
treatment, and no assumptions whatsoever about the source,
propagation or pre-photodetection processing.

Direct measurements of the photon number distribution have been
elusive for technical reasons, but photon counting technology,
namely the visible light photon counter (VLPC)~\cite{Kim99,Tak99},
now provides a means for directly testing nonclassicality for the
(bunched) output field from a PDC as we demonstrate here. Although
criteria for proving nonclassicality of the source via direct
counting have been suggested, such as sub-Poissonian
statistics~\cite{Man79}, observing photon number
oscillations~\cite{Sch87} and deciding if these oscillations are
classical or nonclassical~\cite{Sim97}, or by employing Hillery's
two criteria of comparing the total probabilities of even vs odd
photon numbers being detected~\cite{Hil85}, such criteria are not
necessarily practical, especially for a super-Poissonian
two-photon source.  We introduce a new and practical criterion for
evaluating nonclassicality for weak bunched sources of light and
show that our PDC source \emph{strongly} violates this criterion,
whereas it does not violate Hillery's inequalities. Our violation
of nonclassicality is the first time that nonclassical light has
been unambigously observed by direct photon number detection
without any need for assumptions about the source.

Furthermore the relatively innocuous assumption
that detector inefficiency is due to linear loss processes enables us
to reconstruct from raw data a photon number distribution
that exhibits even-odd photon number oscillations, which strongly
violate Hillery's criteria. Although photon number oscillations have
been reported~\cite{Bre97}, these have in fact been inferred
by performing optical homodyne tomography and thus involve
many assumptions to process the data; in contrast our reconstruction
of photon number oscillations requires no assumptions about the
source and only one assumption about the loss mechanism for detectors.

\subsubsection{Test for non-classical statistics}

With an ideal (i.e., perfect efficiency) photon number detector,
direct observation of PDC output is predicted to exhibit photon number
oscillations~\cite{Sch01}.  For example, with $P_n$ the
probability of observing $n$ photons, an ideal detector should yield
$P_1 = P_3 = 0$ and $P_2 > 0$.  However, an imperfect detector
described by a linear loss model may not directly observe photon
number oscillations, and thus we must establish a criteria for photon
number statistics to be non-classical.  Define $\Gamma$ as the ratio
\begin{equation}
  \label{eq:P_2Ratio}
  \Gamma = \frac{P_2}{P_1 + P_2 + P_3} \, .
\end{equation}
For a perfect detector measuring PDC output, we predict $\Gamma=1$;
detector inefficiencies will lead to a reduced value of this ratio.

We now prove that any semi-classical theory of light which is
constrained to distributions of Poissonian photon statistics cannot
yield states with $\Gamma$ greater than a maximum classical bound
$\Gamma_{\rm classical}$.  For a Poisson photon number distribution
given by $P_n(\bar{n}) = {\rm e}^{-\bar{n}} \frac{\bar{n}^n}{n!}$,
this ratio has a maximum value $\Gamma_{\rm classical} = 3 /
(3+2\sqrt{6}) \simeq 0.379$, saturated by the Poisson distribution
with average photon number $\bar{n} = \sqrt{6}$.  However, one can
show that this optimal value holds not only for a Poisson
distribution, but for any weighted sum of Poisson distributions.
Consider a weighted sum $P_n = \alpha P^{\rm max}_n + (1-\alpha) P'_n$
of two Poisson distributions $P^{\rm max}_n$ and $P'_n$, where $P^{\rm
  max}_n$ has average photon number $\bar{n} = \sqrt{6}$, and $P'_n$
is any other Poisson distribution.  The ratio $\Gamma$ for this
weighted sum is
\begin{equation}
  \label{eq:WeightedSumGamma}
  \Gamma = \frac{\alpha P^{\rm max}_2 + (1-\alpha)
  P'_2}{\alpha(P^{\rm max}_1 + P^{\rm max}_2 + P^{\rm max}_3) +
  (1-\alpha)(P'_1 + P'_2 + P'_3)} \, .
\end{equation}
Because $P^{\rm max}_n$ maximizes $\Gamma$ for any single Poisson
distribution, the mathematical relation
\begin{equation}
  \label{eq:RatioCondition}
  \frac{x'}{y'} < \frac{x}{y} \Rightarrow \frac{\alpha x +
  (1-\alpha)x'}{\alpha y + (1-\alpha)y'} < \frac{x}{y}\, , \quad \forall\
  \alpha<1 \, ,
\end{equation}
proves that $\Gamma \leq 3 / (3+2\sqrt{6})$. Thus, no sum of Poisson
distributions can give rise to a distribution with $\Gamma >
\Gamma_{\rm classical}$

All classical light fields will lead to statistics that can be
expressed as weighted sums of Poisson photon number states.  Thus, the
classical theory of light predicts that the inequality
\begin{equation}
  \label{eq:ClassicalLimit}
  \Gamma \le \frac{3}{3+3\sqrt{6}} ,
\end{equation}
cannot be violated.  In contrast, one expects that light from PDC will
lead to a violation of this condition, which can be demonstrated by
simply measuring $P_1$, $P_2$, and $P_3$.

In the presence of imperfect detection efficiency, loss may
serve to degrade this ratio and a violation of the classical criterion
may not be observed.  Consider a PDC experiment in which the pump is
sufficiently weak that the probability of generating more than one
photon pair is very small.  In this case the ratio in
Eq.~(\ref{eq:P_2Ratio}) is given by $\Gamma = \eta/(2-\eta)$, where
$\eta$ is the detection efficiency.  A violation of the inequality is
not predicted unless $\eta\ge 3/(3+\sqrt{6})\approx 0.55$.

Fortunately, the Visible Light Photon Counter (VLPC) has the
capability to detect photon number states with high quantum
efficiency~\cite{Kim99,Tak99}.  The VLPC has been
shown to have quantum efficiencies approaching $90\%$.  Furthermore,
if more than one photon is incident on the detector surface, the
height of the output electrical pulse is proportional to the number of
incident photons.  This gives us information about the number of
photons that have been detected.

\subsubsection{Observation of non-classical statistics}

\begin{figure}
\centerline{\scalebox{0.6}{\includegraphics{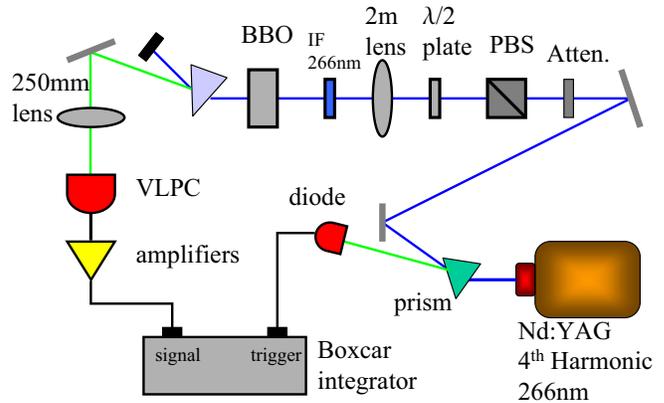}}}
\caption{Schematic of experimental configuration.  The BBO crystal
is tilted for Type I collinear degenerate phase-matching.}
\label{Fig:ExpSetup}
\end{figure}

The experimental setup is shown in Fig.~\ref{Fig:ExpSetup}. We use
the fourth harmonic (266nm) of a Q-Switched ND:YAG laser as a pump
source.  The pulse duration of the laser is 20ns. Using a pulsed
pump allows us to eliminate the detector dark counts
($20,000s^{-1}$) by temporal gating. The laser pumps a BBO crystal
set for collinear degenerate Type I phase matching (optic axis
47.6 degrees from the pump). In this configuration, the
down-converted photons have half the energy of the pump (532nm),
and travel in the same direction. The pump is removed by a prism,
while the down-conversion is focused by a 250mm lens onto the
detector surface.  We also have the option of directly
illuminating the detector with second harmonic light (532nm) from
the laser, which is a classical light source. The detector output
is amplified, and then sent to a gated boxcar integrator, which is
triggered by the laser. The boxcar integrates the pulse over a
20ns window, and the output is sent to an Analog-to-Digital
converter and stored on a computer.

\begin{figure}
\centerline{\scalebox{0.4}{\includegraphics{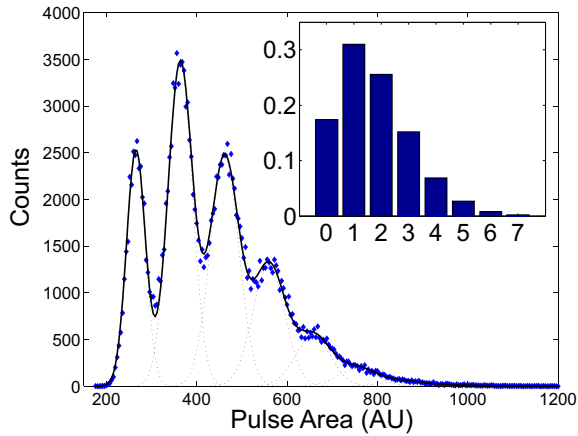}}}
\caption{Pulse area histogram from the boxcar integrator obtained
by illuminating attenuated second harmonic (532nm) light from the
laser onto the detector. The photon number distribution (shown in
the inset) is a Poisson distribution.}\label{Fig:PoissonLight}
\end{figure}

The output of the detector illuminated by light from the second
harmonic of the laser is shown in Fig.~\ref{Fig:PoissonLight}.
The pulse area spectrum features a series of peaks representing
the different photon number state detections.  In the inset we
show the probability distribution, which is calculated by fitting
each peak to a Gaussian function. The area under each Gaussian
curve gives the number of events representing that photon number.
The area of each peak can be normalized by the total area to give
the probability distribution, which is a Poisson distribution as
expected.

\begin{figure}
\centerline{\scalebox{0.6}{\includegraphics{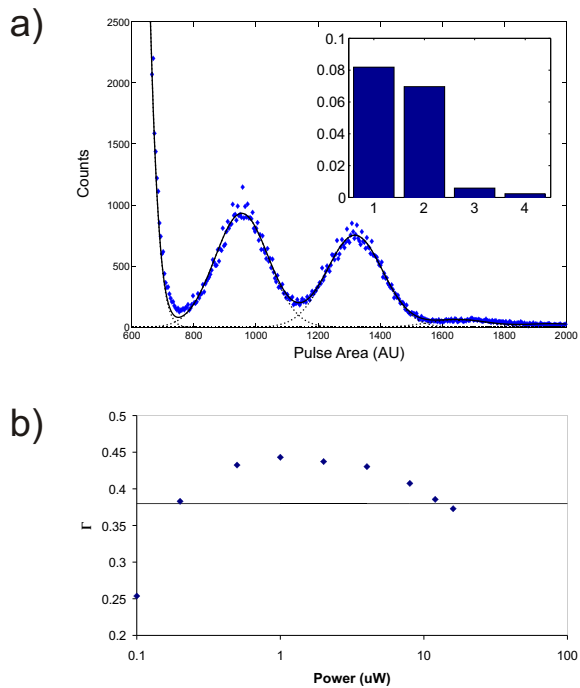}}}
\caption{Observation of non-classical statistics. (a) The pulse
area histogram for detection from the VLPC illuminated by
parametric down-conversion.  The photon number distribution is
shown in the inset.  From this distribution $\Gamma=0.442$, which
violates the classical limit by 40 standard deviations. (b) Plot
of $\Gamma$ as a function of pump power.  At low pumps $\Gamma$
drops due to dark counts.  At high powers it drops due to four
photon events, which enhance the three photon probability after
detection losses.} \label{Fig:PDClight}
\end{figure}

Fig.~\ref{Fig:PDClight}(a) shows the pulse area histogram when the
detector is illuminated by parametric down-conversion, using a
pump power of $1\mu W$. At this weak pump intensity, a single pump
pulse will usually generate zero photons, while a photon pair is
generated with a small probability.  The probability of generating
more than one photon pair is very small. The figure focusses on
the 1, 2, and 3 photon detection peaks, which we will use to
verify non-classical statistics.  The photon number probability
distribution is calculated by fitting each peak to a Gaussian
function.  These areas are normalized by the total area of all the
peaks. The calculated probability distribution is shown in the
inset.  One can see that the probability of 1 and 2 photon
detection is nearly equal, but the probability of 3 photon
detection is nearly zero. These probabilities are $P_1=0.0818$,
$P_2 = 0.0696$, and $P_3=0.0061$, which yields $\Gamma=0.442$,
representing a 40 standard deviation violation of the classical
limit.

The large 1 photon probability is due to losses from the detector and
collection optics.  In the limit of low excitation, the 1 photon and 2
photon probability can be used to calculate the detection efficiency,
given by
\begin{equation} \label{eq:Eta}
  \eta = \frac{2 \frac{P_2}{P_1}}{1 + 2\frac{P_2}{P_1}}.
\end{equation}
From the measurements, it is calculated that the detection
efficiency is 0.67.  Using the measured VLPC quantum efficiency of
$0.85$, the photon collection efficiency is calculated to be
$0.79$.

Fig.~\ref{Fig:PDClight}(b) shows the measured value of $\Gamma$ as a
function of pumping intensity.  The black line represents the
classical limit, which is violated for a large range of pumping
intensities. At high pumping intensities $\Gamma$ begins to drop. This
drop is due to an increase in the two pair creation probability,
which, in the presence of losses, will enhance the 3 photon detection
probability.  The parameter $\Gamma$ also drops at low pumping
intensities.  This drop is attributed to the dark counts of the VLPC.
At low pumping intensities the relative fraction of detection events
originating from dark counts becomes large.  This enhances the 1
photon probability, which reduces the value of $\Gamma$.

\subsubsection{Reconstruction of even-odd oscillations}

\begin{figure*}[t]
\centerline{\scalebox{0.8}{\includegraphics{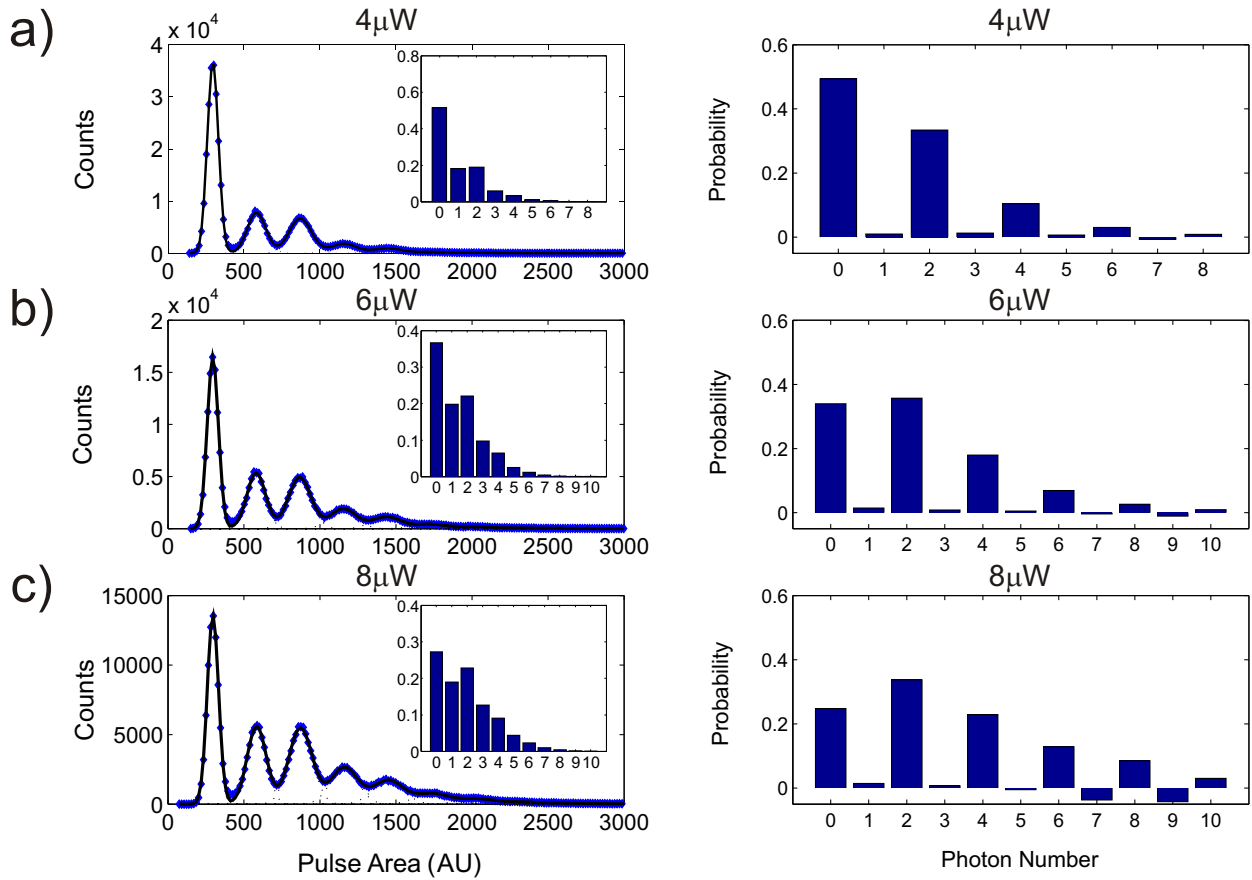}}}
\caption{Pulse area histograms in (a), (b), and (c) are shown for
4, 6, and 8$\mu W$ pumping powers respectively. The probability
distributions are shown in the insets.  By correcting for the
detection efficiency and dark counts we reconstruct the original
photon number distributions shown on the right.  These
distribution show the clear even-odd photon number oscillations
from parametric down-conversion. }\label{Fig:NumOsc}
\end{figure*}

With high detector efficiency, the emitted output of PDC is
predicted to feature even-odd photon number oscillations due to
the two photon nature of the process.  These oscillations lead to
the non-classical statistics discussed in the previous section.
Direct observation of these oscillations using the photon counting
capability of the VLPC would be a remarkable achievement;
unfortunately, direct observation of these oscillations requires
extremely high quantum efficiencies.  Even the relatively high
detection efficiencies of $0.67$ in our experiment are not
predicted to observe this oscillatory behavior.  The requirement
of very high quantum efficiency makes direct observation of the
photon number oscillation extremely difficult in practice.
However, one can make an accurate independent measurement of the
photon detection efficiency, and correct for this effect in the
photon number distribution. This allows the reconstruction of the
original even-odd oscillations of the field.

The detection efficiency can be corrected for as follows. Define $p_i$
as the probability that the photon field contained $i$ photons, and
$f_i$ as the probability that $i$ photons are detected. In the
presence of losses, these two distributions are related by
\begin{equation}
  \label{eq:LossTrans}
  f_i = \sum_{j=i}^{\infty} \binom{j}{i}
  \eta^i \left( 1 - \eta \right)^{j-i} p_i\, .
\end{equation}
In order to calculate $p_i$ from $f_i$, the above transformation
must be inverted.  To perform this inversion, we truncate the
photon number distribution at some photon number $n$, which is
sufficiently large such that $p_{n+1}\approx 0 $ is a good
approximation.  Under this approximation, the initial and final
probability distributions are simply related by a matrix, whose
coefficients are given by Eq.~(\ref{eq:LossTrans}).  The dark
counts of the VLPC can also be accounted for by this matrix, which
can then be inverted to calculate the photon number distribution
of the field.  It is important to emphasize that there are no
fitting parameters in this model.  The only two parameters, the
quantum efficiency and dark counts of the VLPC, are both
independently measured.  Once they are known there is a one-to-one
relationship between the actual and measured photon number
distribution.

Fig.~\ref{Fig:NumOsc} shows the result of the photon number
reconstruction. Three different pumping intensities are shown. For
each pump intensity, the left panel shows the pulse area histogram,
and the inset to the panel shows the calculated photon probability
distribution.  The right panel shows the reconstructed photon number
distribution using the measured quantum efficiency and backgrounds.
The photon number distribution is truncated at 10 photons.  The
reconstructed probabilities demonstrate very clear even-odd
oscillations as predicted from PDC.

At higher photon numbers, it can be seen that the reconstructed
distribution becomes slightly negative.  This erroneous effect is
caused by truncation error.  As the pumping intensity is increased,
the approximation that the photon distribution can be truncated after
10 photons becomes less accurate.  This error manifests itself in the
probabilities becoming slightly negative for the 9 and 7 photon
probability.  This error is worst at the largest pumping intensity of
8$\mu W$, where the truncation approximation is least accurate.  One
could suppress this error by truncating at a higher photon number.
Unfortunately, because of the limited range of the amplifiers and A2D
converters, it is difficult to measure these higher order photon
numbers in practice.  This puts a limit on the pumping power one can
use and still get a good reconstruction.  It is possible that an
improved numerical algorithm over simply putting a cutoff in the
number distribution may overcome some of these practical difficulties.

\subsubsection{Conclusion}

In conclusion, we have directly observed non-classical photon counting
statistics from PDC.  We have shown theoretically that the photon
counting statistics for all classical fields must satisfy the
inequality given in Eq.~(\ref{eq:ClassicalLimit}).  Using the high
quantum efficiency and photon number detection capability of the VLPC
we have experimentally demonstrated violations of this inequality by
light emitted from PDC.  By correcting for the quantum efficiency and
dark counts of the VLPC, we have also succeeded in reconstruction the
even-odd oscillation in the photon number distribution of light
generated by the down-conversion field.

\end{document}